\documentclass[hyper]{JHEP} 

\usepackage{epsfig}

\newcommand\fverb{\setbox\pippobox=\hbox\bgroup\verb}

\newcommand\fverbdo{\egroup\medskip\noindent%

            \fbox{\unhbox\pippobox}\ }

\newcommand\fverbit{\egroup\item[\fbox{\unhbox\pippobox}]}

\newbox\pippobox

\title{Note About String Theory with Deformed Dispersion Relations}
\author{J. Kluso\v{n}\\
Department of
Theoretical Physics and Astrophysics\\
Faculty of Science, Masaryk University\\
Kotl\'{a}\v{r}sk\'{a} 2, 611 37, Brno\\
Czech Republic\\
E-mail: \email{klu@physics.muni.cz}}
\preprint{}

 \abstract{The goal of this short paper is to find Lagrangian for bosonic string  with deformed dispersion relation proposed by
 J.~Magueijo and L.~Smolin in $2004$.  }

\def\be{\begin{equation}}

\def\ee{\end{equation}}

\def\bea{\begin{eqnarray}}

\def\eea{\end{eqnarray}}

\def\bP{\mathbf{P}}

\def\mH{\mathcal{H}}

\def\bB{\mathbf{B}}

\newcommand{\mG}{\mathcal{G}}

\def\pb #1{\left\{#1\right\}}

\begin{document}
\section{Introduction and Summary}\label{first}
Several experiments performed few years ago, as for example high-energy cosmic rays anomalies \cite{Biermann:2002ja,Takeda:1999sg}, suggested that standard
relativistic dispersion relation known from special theory of relativity
\begin{equation}
E^2=p^2+m^2
\end{equation}
may be modified at Planck scales \cite{AmelinoCamelia:2000zs,AmelinoCamelia:2000ge,KowalskiGlikman:2001gp,Magueijo:2002am,Magueijo:2004vv,Magueijo:2001cr}. This modification of dispersion relation
is closely related to so called Double Special Relativity (DSR), for review, see
\cite{KowalskiGlikman:2004qa,KowalskiGlikman:2006vx}. In these theories conventional
dispersion relation $p_M p^M=-m^2$ is generalized to more general one
$p_Mp^M=-m^2+F(\kappa,\dots)$ where $m$ is rest mass and where $\kappa$ is deformation parameter where conventional special relativity dispersion relation is
restored in the limit $\kappa\rightarrow \infty$.

In the past many dispersion relations were proposed that were consistent with $\kappa-$dependent extension of Poincare algebra. In fact, it was also shown how such a dispersion relation could be related to Lagrangian formulation of particle and
deformed symplectic structure \cite{Girelli:2005dc,Mignemi:2003dm,Girelli:2006fw}
\footnote{See also \cite{Deriglazov:2004yr,Pramanik:2013zy,Pramanik:2012fj,Ghosh:2006cb,Ghosh:2006bx}.}.
On the other hand there were no extension of this approach to the case of  string. In fact, string with deformed dispersion relation was proposed
\cite{Magueijo:2004vv} with completely different approach. Explicitly, this proposal was based on modification of the Hamiltonian and spatial diffeomorphism constraint of fundamental string that depend on two functions $f,g$ which depend on total energy of string $\mathbf{P}=\int d\sigma p_0$. It was argued there that such a string could have many interesting properties. The goal of this paper is to investigate this proposal in more details. Especially we are interested in Lagrangian formulation of this theory. To do this we have to deal with the  fact that functions $f$ and $g$ depend on total energy which makes analysis rather complicated. For that reason we introduce two auxiliary global variables $A$ and $B$ and extended Hamiltonian that is equivalent to the original one when we solve appropriate second class constraints. Using this extended Hamiltonian we can now find corresponding Lagrangian that can be even more simpler when we integrate $N^\tau,N^\sigma$ which are Lagrange multipliers of the constraints $\mH_\tau,\mH_\sigma$ respectively. Then we proceed to the most difficult
part of the analysis which is solving equations of motion for $A$ and $B$. We argue that this is possible in principle but it is very difficult. On the other hand we show that for the most interesting case when $f=g$ the Lagrangian does not depend on $f$ and $g$ at all. As a result we find that in case $f=g$ the Lagrangian corresponds to ordinary Nambu-Goto form of the action for bosonic string. In fact, this can be already seen from the Hamiltonian formalism when in case $f=g$ the Hamiltonian reduces to the standard form by redefinition of the Lagrange multipliers $N^\tau$ and $N^\sigma$. In other words, in case $g=f$ there is no modification of the dispersion relation. We show that this is also true in case of point particle with modified Hamiltonian constraint that was also discussed in \cite{Magueijo:2004vv}. In this case the situation is simpler since all modes depend on world-line parameter only but the conclusion is the same. In case when $g=f$ the Lagrangian does not depend on $f$ and $g$ at all and corresponds to ordinary Lagrangian for massive particle.

Let us outline our results. We studied proposal for the string with modified dispersion relation published in \cite{Magueijo:2004vv}. We found corresponding Lagrangian and we argued
that for the most interesting case when $f=g$ Lagrangian does not depend on $f$ abd $g$  and
hence reduces to ordinary Nambu-Goto action which certainly cannot lead to string with modified dispersion relation. We mean that this is very interesting result. Of course, this result does not show that it is not possible to construct string with modified dispersion relation. However we mean that we should rather follow point particle analysis that can be found for example in
 \cite{AmelinoCamelia:2000zs,AmelinoCamelia:2000ge,KowalskiGlikman:2001gp,Magueijo:2002am,Magueijo:2004vv,Magueijo:2001cr}. Explicitly, we should start with modified Hamiltonian constraint and try to construct corresponding Lagrangian for bosonic string that will be diffeomorphism invariant. This problem is currently under investigation.

\section{String with Deformed Dispersion Relations}\label{second}
Following \cite{Magueijo:2004vv} we consider action for bosonic string in the form
\begin{equation}
S=\int d\tau d\sigma (p_M\partial_\tau x^M-N^\tau \mH_\tau-N^\sigma \mH_\sigma) \ ,
\end{equation}
where the Hamiltonian and spatial diffeomorphism constraints
$\mH_\tau,\mH_\sigma$ are defined as \cite{Magueijo:2004vv}
\begin{eqnarray}
\mH_\tau=\frac{f}{2T}p_M g^{MN}p_N
+\frac{Tg}{2}g_{MN}\partial_\sigma x^M\partial_\sigma x^N \ , \quad
\mH_\sigma=\sqrt{fg}p_M\partial_\sigma x^M \ ,
\end{eqnarray}
where $f,g$ are functions that depend on total energy
\cite{Magueijo:2004vv}
\begin{equation}\label{bP}
\bP_0=\int d\sigma p_0 \ .
\end{equation}
Further, $g_{MN},M,N=0,\dots,25$ is background metric with inverse
$g^{MN}$ and $x^M(\tau,\sigma)$ parameterize an embedding of
fundamental string into target space-time and $p_M$ are conjugate
momenta. Finally $T$ is string tension.  Note that the world-sheet
of string is parameterized with $\tau,\sigma$ and we presume that
the string is finite with the length $l$ so that $\sigma\in (0,l)$.

To proceed further we will replace (\ref{bP}) with following
covariant prescription
\begin{equation}\label{global}
K^M\int d\sigma p_M
 \ ,
 \end{equation}
 where $K^M$ is constant vector. This vector cannot depend on space-time coordinates
 since in the opposite case it would depend on $x^K(\sigma)$ and hence  (\ref{global}) were not global. Further, we can easily restore (\ref{bP}) when
 we chose $K^M=\delta^M_0$.

 Since functions $f$ and $g$ depend on global quantity it is convenient to introduce auxiliary world-volume variables $A(\tau),B(\tau)$ that depend on $\tau$ only. To do this let us consider following Hamiltonian
 \begin{eqnarray}\label{Hext}
H=\int d\sigma(N^\tau \mH_\tau(A)+N^\sigma \mH_\sigma(A))+B(K^M\int d\sigma
p_M-A) +v_A p_A+v_Bp_B\ , \nonumber \\
\end{eqnarray}
where
\begin{equation}
\mH_\tau(A)=\frac{f(A)}{2T}p_M g^{MN}p_N
+\frac{Tg(A)}{2}g_{MN}\partial_\sigma x^M\partial_\sigma x^N \ , \quad
\mH_\sigma(A)=\sqrt{fg}p_M\partial_\sigma x^M \ .
\end{equation}
As (\ref{Hext})  suggests momenta conjugate to $A$ and $B$ respectively,   $p_A(\tau)\approx 0\ , p_B(\tau)\approx 0$ are primary constraints together with $p_\tau\approx 0,p_\sigma\approx 0$ which are momenta conjugate to $N^\tau$ and $N^\sigma$. Now requirement of the preservation of primary constraints implies following secondary constraints
\begin{eqnarray}
& &\partial_\tau p_A=\pb{p_A,H}=B-\int d\sigma N^\tau\frac{\delta \mH_\tau}{\delta A}
-\int d\sigma N^\sigma \frac{\delta \mH_\sigma}{\delta A}\equiv \mG_A\approx 0 \ ,
\nonumber \\
&&\partial_\tau p_B=\pb{p_B,H}=-K^M\int d\sigma p_M+A\equiv \mG_B\approx 0 \ ,
\nonumber \\
& &\partial_\tau p_N=\pb{p_N,H}=-\mH_\tau \approx 0 \ , \nonumber \\
& &\partial_\tau p_\sigma=\pb{p_\sigma,H}=-\mH_\sigma \approx 0 \ .
\nonumber \\
\end{eqnarray}
Clearly $p_A,p_B,\mG_A,\mG_B$ are second class constraints that can be explicitly solved for $A$ and $B$. In fact, from $\mG_B$ we express $A$ as $A=K^M\int d\sigma p_M$ and inserting it to $\mH_\tau,\mH_\sigma$ we get original Hamiltonian. Then from $\mG_A$ we could express $B$ as functions of remaining dynamical variables however this is not very important due to the fact that $B$ is multiplied by $\mG_B$ in Hamiltonian  and hence its contribution vanishes. In other words we have shown equivalence of the original and extended Hamiltonian. It is also clear that $\mH_\tau(A),\mH_\sigma(A)$ are first class constraints since they have the same structure as in case of  ordinary bosonic string.

Now we are ready to find Lagrangian form of this theory. Using (\ref{Hext}) we obtain following equation of motion for  $x^M$
\begin{equation}
\partial_\tau x^M=\pb{x^M,H}=
N^\tau \frac{f}{T}g^{MN}p_N+N^\sigma\sqrt{fg}\partial_\sigma x^M+BK^M
\end{equation}
and hence
\begin{equation}
\partial_\tau x^M-N^\sigma \sqrt{fg}\partial_\sigma x^M-BK^M=
N^\tau \frac{f}{T}g^{MN}p_N \ .
\end{equation}
Using this result we obtain  Lagrangian  as
\begin{eqnarray}\label{L}
& &L=\int d\sigma(p_M\partial_\tau x^M)-H=\int d\sigma\left(\frac{f}{2T}p_M g^{MN}p_N
+\frac{N^\tau Tg}{2}g_{MN}\partial_\sigma x^M
\partial_\sigma x^N+BA\right)=\nonumber \\
& &=\int d\sigma [\frac{T}{2fN^\tau}(g_{\tau\tau}
-2N^\sigma \sqrt{fg}g_{\tau\sigma}+(N^\sigma)^2fg g_{\sigma\sigma}-2\partial_\tau x^M
g_{MN}K^N B+2N^\sigma \sqrt{fg}\partial_\sigma x^M g_{MN}K^N B\nonumber \\
& &+B^2K^Mg_{MN}K^N)
-N^\tau\frac{Tg}{2}g_{\sigma\sigma}+BA]  \ , \nonumber \\
\end{eqnarray}
where $g_{\alpha\beta}=g_{MN}\partial_\alpha x^M\partial_\beta x^N$ and
where we again stress that  $A,B$ depend on $\tau$ only.

As the next step we solve equations of motion for $N^\sigma$ and $N^\tau$ that follow from (\ref{L}). We start with the equation of motion for $N^\sigma$ that has the form
\begin{equation}
-\sqrt{fg}g_{\tau\sigma}+\sqrt{fg}\partial_\sigma x^M g_{MN}K^N B+
N^\sigma fg g_{\sigma\sigma}=0
\ .
\end{equation}
It can be solved for $N^\sigma$ and we obtain
\begin{equation}\label{Nsigmasol}
N^\sigma=\frac{1}{\sqrt{fg}g_{\sigma\sigma}}
    (g_{\tau\sigma}-\partial_\sigma x^M g_{MN}K^N B) \ .
\end{equation}
Further, equation of motion for $N^\tau$ that follow from (\ref{L}) has the form
\begin{eqnarray}
& &-\frac{T}{2f(N^\tau)^2}
\left(g_{\tau\tau}
-2N^\sigma \sqrt{fg}g_{\tau\sigma}+(N^\sigma)^2fg g_{\sigma\sigma}-2\partial_\tau x^M
g_{MN}K^N B+\right. \nonumber \\
& &+\left.
2N^\sigma \sqrt{fg}\partial_\sigma x^M g_{MN}K^N B
+B^2K^Mg_{MN}K^N\right)
-\frac{Tg}{2}g_{\sigma\sigma}=0 \ . \nonumber \\
\nonumber \\
\end{eqnarray}
Inserting (\ref{Nsigmasol}) into equation above we can solve if for $N^\tau$ with the result \begin{eqnarray}\label{Ntausol}
& &(N^\tau)^2=-\frac{1}{fg g_{\sigma\sigma}^2}\left
[\det g_{\alpha\beta}+2(g_{\tau\sigma}\partial_\sigma x^M g_{MN}K^N-
\partial_\tau x^M g_{MN}K^N)B+\right.\nonumber \\
&&\left.(-(\partial_\sigma x^M g_{MN}K^N)^2+K^Mg_{MN}K^N)B^2\right] \ . \nonumber \\
\end{eqnarray}
Finally inserting (\ref{Nsigmasol}) and (\ref{Ntausol}) into (\ref{L}) we find that the Lagrangian has the form
\begin{eqnarray}\label{Lfinal}
& &L=-\int d\sigma [\sqrt{\frac{g}{f}}\left[-
(\det g_{\alpha\beta}+2(g_{\tau\sigma}\partial_\sigma x^M g_{MN}K^N-
\partial_\tau x^M g_{MN}K^N)B+\right.\nonumber \\
& & \left. ((\partial_\sigma x^M g_{MN}K^N)^2+K^Mg_{MN}K^N)B^2)\right]^{1/2}+BA] \ .
\end{eqnarray}
As the last step  we try to solve equations of motion for $A$ and $B$ that follow from (\ref{Lfinal}). First of all the  equation of motion for $A$ has the form
\begin{equation}\label{deltaA}
\frac{\delta}{\delta A}\sqrt{\frac{g}{f}}
\int d\sigma
\sqrt{\bB}+lB=0 \ ,
\end{equation}
where $l$ is the length of the string, and where we defined $\bB$ as
\begin{eqnarray}
& &\bB=
\det g_{\alpha\beta}+2(g_{\tau\sigma}\partial_\sigma x^M g_{MN}K^N-
\partial_\tau x^M g_{MN}K^N)B+\nonumber \\
& &+((\partial_\sigma x^M g_{MN}K^N)^2+K^Mg_{MN}K^N)B^2 \ .
\end{eqnarray}
 The equation (\ref{deltaA}) can be solved  for $A$, at least in principle when we introduce $\Psi$ as function inverse to $\frac{\delta}{\delta A}\sqrt{\frac{g}{f}}$. Then we have
\begin{equation}\label{APsi}
A=-\Psi \left[\frac{lB}{\int d\sigma \sqrt{\bB}}\right]  \ .
\end{equation}
Further,  equation of motion for $B$ has the form
\begin{eqnarray}
& &\int d\sigma\left[\sqrt{\frac{g}{f}}(
(g_{\tau\sigma}\partial_\sigma x^M g_{MK}K^K-\partial_\tau x^M g_{MN}
K^N)+\right.\nonumber \\
& &+\left.((\partial_\sigma x^M g_{MN}K^N)^2+K^Mg_{MN}K^N)B)\frac{1}{\sqrt{\bB}}+A\right]=0 \ .
\nonumber \\
\end{eqnarray}
If we now insert  (\ref{APsi}) into equation above it can  be solved for $B$ at least in principle. In other words we obtain $A=\Sigma(p_M,x^M)$ and
$B=\Omega(p_M,x^M)$.
Inserting back to the Lagrangian density we obtain its final form which is very complicated. On the other hand it was argued in \cite{Magueijo:2004vv} that the only meaningful situation occurs when $f=g$.  Then however looking on the Lagrangian
(\ref{Lfinal}) we see that it does not depend on $f=g$ at all. Moreover, the equation of motion for $A$ implies that $B=0$. Further, equation of motion for $B$ can be solved for $A$
however due to the fact that $A$ multiplies $B$  we see that it does not contribute to the final Lagrangian at all. In summary we find that the Lagrangian (\ref{Lfinal}) has the form
\begin{eqnarray}\label{Lfinalf}
L=-\int d\sigma \sqrt{-\det g_{\alpha\beta}}
\end{eqnarray}
which is standard Nambu-Goto form of Lagrangian for bosonic string. In other words the case when $g=f$ corresponds to ordinary bosonic string. In fact, this can be already seen from the form of  Hamiltonian that for $f=g$ has the form
\begin{equation}
H=\int d\sigma \left(N^\tau f\left(
\frac{1}{2T}p_M g^{MN}p_N
+\frac{T}{2}g_{MN}\partial_\sigma x^M\partial_\sigma x^N\right)+N^\sigma f p_M\partial_\sigma x^M\right) \ .
\end{equation}
Performing redefinition of Lagrange multipliers as $N^\tau f\rightarrow N^\tau, N^\sigma f\rightarrow N^\sigma$ we obtain again Hamiltonian for bosonic string. Such a redefinition can be always done since $N^\tau,N^\sigma$ are parameters of world-sheet diffeomorphism. In other words the case when $f=g$ cannot lead to modified dispersion relation.

In order to gain more insight into this problem  we will further discuss  the case of point particle with modified dispersion relation
\cite{Magueijo:2001cr}.
\section{Point Particle with modified Hamiltonian Constraint}\label{third}
Following \cite{Magueijo:2001cr} we consider point particle whose dynamics is governed by the action
\begin{equation}
S=\int d\tau (p_M\dot{x}^M-N\mH) \ ,
\end{equation}
where the Hamiltonian constraint has the form
\begin{equation}
\mH=\frac{f(K^Mp_M)}{2}g^{MN}p_Mp_N+\frac{1}{2}g(K^M p_M)(m^2+V(x)) \ .
\end{equation}
We again try to find Lagrangian form of this action. Note that now
all modes depend on $\tau$ only and we do not need to distinguish
between global and local variables from world-sheet point of view as
we did in previous section.

To begin with  we introduce auxiliary
modes $A$ and $B$ so that we can write extended Hamiltonian constraint in the form
\begin{equation}
N\mH_E=N(\frac{f(A)}{2}g^{MN}p_Mp_N+\frac{1}{2}g(A)(m^2+V(x)))+B(K^Mp_M-A)+u^Ap_A+u^Bp_B \ ,
\end{equation}
where $u^A,u^B$ are Lagrange multipliers that ensure that $p_A\approx 0 , p_B\approx 0$ are primary constraints. Then the requirement of the preservation of these primary constraints implies secondary constraints
\begin{eqnarray}
& &\partial_\tau p_A=\pb{p_A,H}=-\frac{N}{2}f'(A)p_Mg^{MN}p_N-Ng'(A)(m^2+V(x))+B\equiv
\mG_A\approx 0 \ ,
\nonumber \\
& &\partial_\tau p_B=\pb{p_B,H}=-(K^Mp_M-A)\equiv -\mG_B\approx 0 \ .
\nonumber \\
\end{eqnarray}
Clearly $\mG_A,\mG_B,p_A,p_B$ are second class constraints that vanish strongly. Solving $\mG_B$ for $A$ we get original action while solving $\mG_A$ for $B$ we can express $B$ as function of canonical variables however this is not necessary since $B$ appears in the action with the combination with $\mG_B$. As a result we obtain original action.

Using this extended Hamiltonian we now obtain following equations of motion for
 $p_M,x^M$
\begin{eqnarray}
\dot{x}^M=Nf(A)g^{MN}p_N+BK^M \ , \quad
\dot{p}_M=-\frac{N}{2}\partial_M V(x)-N\frac{f(A)}{2}\partial_M g^{KL}p_Kp_L \ .
\nonumber \\
\end{eqnarray}

Then the  Lagrangian is equal to
 \begin{eqnarray}\label{LE}
& &L=p_M\dot{x}^M-N\mH_E=
\nonumber \\
& &=
\frac{1}{Nf}(\dot{x}^M-BK^M)g_{MN}(\dot{x}^N-BK^N)-\frac{N}{2}g(A)(m^2+V)+BA \ .
\nonumber \\
\end{eqnarray}
Now we again solve equation of motion for $N$ that follow from
(\ref{LE})
\begin{equation}
-\frac{1}{N^2f}(\dot{x}^M-BK^M)g_{MN}(\dot{x}^N-BK^N)-\frac{1}{2}g(A)(m^2+V)
=0
\end{equation}
that can be solved for $N$ with the result
\begin{equation}
N=\sqrt{-\frac{(\dot{x}^M-BK^M)g_{MN}(\dot{x}^N-BK^N)}{fg(m^2+V)}} \ .
\end{equation}
Inserting this result back to (\ref{LE}) we obtain that Lagrangian has the form
\begin{equation}\label{LEe}
L=-\sqrt{\frac{g}{f}}
\sqrt{-(\dot{x}^M-BK^M)g_{MN}(\dot{x}^N-BK^N)}
\sqrt{m^2+V}+BA \ .
\end{equation}
Finally we study equations of motion for $A$ and $B$ that follow from (\ref{LEe})
The equation of motion for $A$ has the form
\begin{equation}
-\Phi'\sqrt{-(\dot{x}^M-BK^M)g_{MN}(\dot{x}^N-BK^N)}
\sqrt{m^2+V}+B=0 \ ,
\end{equation}
where we defined $\Phi=\sqrt{\frac{g}{f}}$ and where
 $\Phi'=\frac{d\Phi}{dA}$. The equations of motion for $B$ has the form
\begin{equation}
-\Phi \frac{K^Mg_{MN}(\dot{x}^N-BK^N)}{
\sqrt{-(\dot{x}^M-BK^M)g_{MN}(\dot{x}^N-BK^N)}}
\sqrt{m^2+V}
+A=0 \ .
\end{equation}
It is very difficult to solve these equations in the full generality. However for $
\Phi=1$ we obtain that $B=0$  while from the equation of motion for $B$ we obtain that $A$ is equal to
\begin{equation}
A=\frac{K^M g_{MN}\dot{x}^N}{\sqrt{-\dot{x}^Mg_{MN}\dot{x}^N}}\sqrt{m^2+V} \ .
\end{equation}
 Again this result is not important since $A$ is multiplied by $B$ in Lagrangian and as we know equation of motion for $A$ implies that $B=0$. As a result we obtain final form of Lagrangian
\begin{equation}
L=-\sqrt{-\dot{x}^M g_{MN}\dot{x}^N}\sqrt{m^2+V} \ .
\end{equation}
We  see that the preferred case $f=g$ corresponds to the ordinary massive particle without any modification of the dispersion relation that confirms result derived in
the previous section.

{\bf Acknowledgement:}
\\
This work
is supported by the grant “Integrable Deformations”
(GA20-04800S) from the Czech Science Foundation
(GACR).

\end{document}